\documentclass[sigconf, auth]{acmart}
\usepackage{graphicx}
\usepackage{balance}
\begin{document}

\title{How are Software Repositories Mined? \\A Systematic Literature Review of Workflows, Methodologies, Reproducibility, and Tools}


\author{Adam Tutko}
\email{atutko@vols.utk.edu}
\orcid{0003-2412-5409}
\affiliation{%
  \institution{University of Tennessee}
  \city{Knoxville}
  \state{TN}
  \postcode{37996}
}
\author{Audris Mockus}
\email{audris@utk.edu}
\orcid{0002-7987-7598}
\affiliation{%
  \institution{University of Tennessee}
  \city{Knoxville}
  \state{TN}
  \postcode{37996}
}
\author{Austin Z. Henley}
\email{azh@utk.edu}
\orcid{0003-1069-2795}
\affiliation{%
  \institution{University of Tennessee}
  \city{Knoxville}
  \state{TN}
  \postcode{37996}
}

\begin{abstract} \label{0}
With the advent of open source software, a veritable treasure trove of previously proprietary software development data was made available. This opened the field of empirical software engineering research to anyone in academia. Data that is mined from software projects, however, requires extensive processing and needs to be handled with utmost care to ensure valid conclusions. Since the software development practices and tools have changed over two decades, we aim to understand the state-of-the-art research workflows and to highlight potential challenges. We employ a systematic literature review by sampling over one thousand papers from leading conferences and by analyzing the 286 most relevant papers from the perspective of data workflows, methodologies, reproducibility, and tools. We found that an important part of the research workflow involving dataset selection was particularly problematic, which raises questions about the generality of the results in existing literature. Furthermore, we found a considerable number of papers provide little or no reproducibility instructions---a substantial deficiency for a data-intensive field. In fact, 33\% of papers provide no information on how their data was retrieved. Based on these findings, we propose ways to address these shortcomings via existing tools and also provide recommendations to improve research workflows and the reproducibility of research.    
\end{abstract}

\maketitle

\section{Introduction and Motivation} \label{sec-intro}


The public nature of open source software and the embrace of tools
such as version control and issue tracking by software developers
has created volumes of data that, using suitable methods, can be
exploited to construct precise measures and models of software
development. The process of obtaining and transforming data from
such tools is the main objective of the field of mining software
repositories. The field has expanded and matured over the last two
decades in many ways. For example, the version control and issue
tracking tools have evolved significantly since their creation. 
Alongside this, many other tools have been created that may produce traces that shed light on software development. 

The open source software -- the primary
subject of research -- has experienced tremendous growth and now
occupies a much more central role in the overall software
ecosystem. The software development process is now more attuned to
dealing not just with issues within a project, but also with
upstream and, often, downstream dependencies. Such dependencies are typically managed by
entirely different teams and organizations and this situation adds to the convolution of the network. 
Such changes to the ecosystem raise questions about whether the primary workflows, developed decades ago, still
serve the needs of researchers. Apart from changed
data sources, application domain, and software development practices,
the research process itself has changed as well. For example,
reproducibility is considered a much more important issue than
even a decade ago. A lot more rigor is expected of the analysis.
This is especially true as understanding of various pitfalls related to
development begin to support tool specific data (see, e.g.,~\cite{M14}).  

Systematic literature reviews~\cite{kitchenham2009systematic} are
best suited to address our research questions: i.e., to summarize
the state of the art and to elucidate remaining challenges in the
field of mining software repositories. Since this field relies on processing large amounts of
data, our primary focus in the systematic literature review was to
identify and summarize the workflows used, the choice of datasets
and tools employed, and the reproducibility of the findings. We
especially focus on methods used to identify what software projects
are selected for the analysis as such selection (or sampling)
greatly affects the generalizability of the results.


The mining software repositories field is a field of research focused on the retrieval and analysis of data. The field has grown and changed significantly in the last 15 years. Mining software repositories research was originally focused on CSV repositories. Later, the focus shifted to SVN repositories and in recent years the focus has mostly shifted to Git repositories or websites such as Stack Overflow. Due to the evolution of this field, researchers and their methods have had to change alongside it.

Finding ways to do the research in intelligent and efficient ways is a big step of the process. In pursuit of this, researchers have dedicated many hours to the creation of tools and datasets to make the data mining process easier. Indeed, many of the papers published in mining software repositories relevant conferences can be categorized as a tool or dataset designed to make this process easier. However, despite the prevalence of these tools, the frequency of categorical usage of these tools is not obvious. Given the list of tools that have been published, determining how frequently they are used could provide valuable insight into the research process. 

To garner a better understanding of the state of the practice of mining software repositories research, we performed a systematic literature review of three mining software repositories adjacent conferences. With the intent of applying this research to current and relevant data, we restricted our literature review from 2018 to 2020. The three conferences included in the literature review are the International Conference on Software Engineering (ICSE), European Software Engineering Conference and Symposium on the Foundations of Software Engineering (ESEC/FSE), and The Mining Software Repositories (MSR) conference. In doing so, we aim to answer the following research questions:

\begin{itemize}

  \item RQ1: What workflows do researchers use to mine software repositories?
  
  \item RQ2: What methodological decisions do researchers make and report in mining software repository papers?
  
  \item RQ3: What tools are used to mine software repositories?
  
  \item RQ4: Are the data sampling and data retrieval aspects of this field of research reproducible? 
\end{itemize}

A total of 1109 research papers were looked at in this literature review and a total of 286 papers were collected and categorized as relevant from these three conferences. A pre-selection criteria was applied to the papers and then each was manually analyzed by one of the authors. From the data discerned in the literature review, the team determined the general statistics of tool usage when performing mining software repositories research, the general workflow of researchers, and analyzed the replicability and reproducibility of the current research process. 
This paper contributes:
\begin{itemize}

  \item A systematic literature review considering 286 recent papers to investigate the state of the practice. All papers analyzed and the data used to perform the evaluation is listed on our online appendix\footnote{https://zenodo.org/record/5274208\#.YSlMs7RKjs0}.
  
  \item A general look into the workflow of researchers when performing sampling or retrieval and a general understanding of the state of the art of the research.
  
  \item A categorical list of the most popular tools and datasets in the field and statistics of dataset re-use.
  
  \item A look into the reproducibility of mining software repositories research and a suggested format to make for easier reproduction of published works in the future.
\end{itemize}

\section{Background and Related Work} \label{sec-background}

The early empirical work based on software development data defined
methodology for workflows that processed version control and issue
tracking data but was primarily focused on the software engineering
lessons, see, e.g.,~\cite{Mockus00,MFH00}. Soon thereafter, some
aspects of the methodology were considered as stand-alone
contributions,
e.g.,~\cite{M02,GM03,Zimmermann,MTutorial06,Changes07}. Once the
field was maturing, specialized methodological contributions
started to appear, for example, challenges of analyzing data from
git repositories~\cite{msrgit} or more generally~\cite{M14}.
A look at data sources, what and why software repository mining was
conducted for, methodology used and ways to assess quality were
considered in the literature survey by Kagdi et al~\cite{Kagdi2007JSME}. 
More recently, researchers tried to identify problems in existing
literature cause by data leakage (when prediction is made using data
not available at the time of prediction)~\cite{tu2018careful} and to
assess the impact of time stamp anomalies in git data on
existing results~\cite{flint2021escaping}.




In the field of mining software repositories research, researchers have a lot of potential sources of data acquisition due to the long list of Git based software hosting platforms. However, the vast majority of papers that perform this kind of work choose to do so by relying on the hosting site GitHub. This is likely due to the fact that it is the number one repository host in the world. GitHub has over 65 Million registered users and more than 200 million repositories~\footnote{https://github.com/about}. 
In comparison, GitLab, the second most popular platform, has a total of 30 million registered users\footnote{https://about.gitlab.com/company/} with no clear number of hosted repositories. 
Considering the disparity in data availability, it is no wonder researchers tend to rely on GitHub as their source of data. In fact, there are a list of tools and datasets meant to help with the mining software repositories process and many of them rely on GitHub as their sole source of information.

Tools like GHTorrent \cite{GHtorrentGousi13}, the World of Code \cite{WOCma2019world}, Boa \cite{BOAdyer2015},  all rely on GitHub as a resource at least partially. Furthermore, all of these tools provide a precompiled dataset for researchers to gather their data. All of these tools were designed to help ease research in the software engineering process by precompiling much of the relevant data into one place. GHTorrent is a tool that monitors the GitHub public event timeline and processes the data that is posted there into a set of databases for users to access. The World of Code is a tool that attempts to document the entire Free/Libre Open Source Software (FLOSS) ecosystem by monitoring a long list of hosting sites for previously unfound data. This data comes from many of the popular hosting sites ( e.g. GitHub and GitLab), but attempts to also include more obscure domains through a list of heuristic discovery tools. Boa is a similar tool that precompiles a set of relavent open source projects into a dataset that can be later analyzed. The Boa dataset is highly language specific as the projects included are restricted to Java and Python. 
Similarly, SOTorrent \cite{DBLP:conf/msr/BaltesDT008} is a tool that was built to mirror GHTorrent's purpose but for Stack Overflow's publicly posted Stack Overflow Dumps\footnote{https://archive.org/details/stackexchange}.

These are not the only types of tools though. Tools like PyDriller \cite{PyDrillerSpadini2018} or the GitHub REST API were designed to assist with the mining process by automating steps within the collection process. PyDriller allows users to easily extract data from git repositories like commits, developers, source code, etc. The GitHub REST API allows users to search and retrieve data within GitHub based on some prebuilt flags. Such tools provide an excellent way to perform comprehensive research in the field without the needing to perform any personal mining of data. Alongside this, there are tools meant to help with the data selection process. GHS (GitHub Search) \cite{dabic2021sampling}, a relatively recent tool, was created to help researchers sample their data solely through this system. Like other tools, it attempts to collect a list of data it believes is relevant to the field and provide it to users in precompiled format.  

\section{Methodology} \label{sec-method}

When performing mining software repositories research, there are
many steps and multiple ways to perform the work.
As described in~\cite{Changes07}, analysis of software repositories
can include an array of possibilities. Frequently, it includes
understanding the practices of how the development support
systems are used in a project, retrieval of raw data from
development support systems, augmentation of raw data with
quantities that are not directly recorded in these systems,
producing meaningful measures by operationalizing quantities that
are the subject of research, and modeling the resulting measures
using appropriate statistical or machine learning tools. The
validation is applied after each step and, if any issues are found,
a new iteration of research is started. A high-level description of this process is further described in ~\cite{MTutorial06}. 


However, prior work on describing mining software repositories does not represent the full
contemporary workflow. Specifically, ~\cite{MTutorial06}'s work implies that a single project is
analyzed, but many contemporary workflows incorporate multiple
projects selected to represent a certain area. 
The general trend towards more external and internal
validity and reproducibility in software engineering research must
have had a significant influence on the basic workflow and, possibly, added or refined
a number of steps in it.
Having a clear understanding of the current research workflow should
not only help evaluate research contributions but also help 
improve it. 
To achieve this, we performed a systematic literature review of papers from three research conferences. 
Papers were pre-selected based on a relevance criterion then further hand-selected to include only relevant papers.
The papers were then analyzed to garner information on the process followed to perform the work. 

To begin our literature review we started by restricting the papers being reviewed to three main conferences. The conference most relevant to the work is the Mining Software Repositories (MSR) conference. However, to make sure the data generalizes across research, we included premier conferences that tend to publish mining software repository research. These are the International Conference on Software Engineering (ICSE) and the European Software Engineering Conference and Symposium on the Foundations of Software Engineering (ESEC/FSE), which are internationally renowned conference in the general field of software engineering.

To begin selecting the relevant papers, all of the papers from the conferences hosted in 2018, 2019, and 2020 were downloaded. An initial pass was performed on the papers to identify keywords for inclusion criteria. 
The keyword identification was done through text search for the words GitHub or Stack Overflow or Stack Exchange as well as by reading the abstract and introduction to note papers that mention relevant tools. The inclusion criteria we used is:
\begin{itemize}
    \item Papers that mention data collection/sampling from GitHub,
      Stack Exchange, or Stack Overflow
    \item Papers that mention GHTorrent, World of Code,
      Software Heritage, SOTorrent, or Boa. 
\end{itemize}

To expedite this process, papers that did not include the words
GitHub, Stack Overflow, Stack Exchange, or one of the tools reliant
on those sources were summarily excluded.  


After the first pre-selection, one author manually analyzed the remaining
papers to further remove papers that mention these things
tangentially (e.g. in the related work section). The sections analyzed in each paper
were the methodology section as well as any experimental setup section. 
Not all papers had a section specifically denoted Methodology.
Thus, papers were scanned for section names that had similar meaning to methodology or workflow, 
and those sections were analyzed. After paper
selection, the same author analyzed each paper looking for
information on how the researcher(s) in each paper performed the MSR
process. Specifically, the author looked for information on each
paper's methodology and workflow. This included: 
\begin{itemize}

\item Determine which tools were used to conduct the research (if
  any) and if any custom tools were created. 

\item Determine methods for sampling or selecting the data. 

\item Determine if the paper re-used MSR datasets and if it published any.

\item Determine if a replication package was included.

\item Determine the steps researchers followed in their work and in what order
  so that a general understanding of their workflow could be understood 
  (e.g. in what way did they sample the data and did they collect it).

\end{itemize}



Separating tools and datasets was a challenge as both were used for similar
purposes. Thus, for our
literature review, we classify tools as resources that allow direct
access to/retrieval of data and classify datasets as resources
that do not allow for direct data retrieval, but provide a selection
of data. An example of the difference between these classifications
alongside our results for these tools/datasets are listed in
Section~\ref{sec-results-methodologies} and
Section~\ref{sec-results-tools}. 

In a similar avenue of focus, we attempted to determine how
frequently researchers were required to write custom scripts to
conduct their research. To ensure a deeper look into the
workflow, script creation was evaluated for each step of the
the workflow. This was done to help determine in which portion of the
workflow researchers most frequently had to write scripts to
determine where the pain points in the process were most
frequent. This part of the analysis is also detailed in
Section~\ref{sec-results-tools}. 

In an attempt to garner an understanding on how researchers select
their data, we also analyzed how researchers performed the sampling
of data. Sampling is a common practice in scientific research and is
done when it is impractical to do research on the entire population.
Sampling entails selecting a few cases in such a way as to represent the
entire population from a certain perspective. In other words, the
findings from the selected sample should be the same as findings if
the research was done on the entire population. The sheer amount
of open source data available tends to require sampling but it is
not always readily apparent how to perform the sampling. Thus,
figuring out the most common sampling practices might provide
insight into the process researchers follow and suggest
improvements. The results of this analysis are detailed in
Section~\ref{sec-results-methodologies}.

Research involving complex data that requires extensive processing
tends to be impossible to replicate without properly prepared
replication packages that include data and programs used to process
it. 
We, therefore, check if researchers claimed to have a replication package. Alongside this, the specific data contained within the package was tracked as well. Things like the availability of the dataset used, availability of personal scripts, and access to any resources used to garner their results were tracked. This is detailed in Section~\ref{sec-results-replication}.

Finally, we aimed to develop a generalized workflow that could describe the process taken to mine software repositories in our literature review.
To do so, we read the methodology section in approximately half of the papers selected for our literature review and inductively developed a workflow that could apply to all of them.
We then validated our generalized workflow by checking to see if the remaining half of the papers from our literature review were accurately described by our generalized workflow.
This mining software repositories research workflow can be
approximately categorized into three steps: sampling, filtering, and retrieval. These steps are further detailed in Section~\ref{sec-results-workflows}.

\section{Results} \label{sec-results}

To address our research questions, we analyzed the selected papers in terms of workflows, methodologies, reproducibility, and tools. For RQ1, we analyze the workflow of researchers performing mining software repositories analysis. For RQ2, we gathered the descriptions of the methodological choices researchers made when performing their work. Specifically, we focus on how researchers select the data they were interested in. For RQ3, we describe the tools most frequently used in the research. The purpose of this was to elucidate the existence of retrieval bias or reliance. Finally, for RQ4, we provide statistics on the reproducibility of researchers' work. The remainder of this section describes the results to address each research question in detail.

\subsection{RQ1 Results: Workflows} \label{sec-results-workflows}

Mining software repositories has changed over time
and, more generally, researchers tend to differ in their choices on
how to do their work. In attempt to analyze this step in the
research process, we compiled the workflow steps and choices of each
research paper based on the results from the survey to determine
what steps were taken most frequently. This was done in attempt to
garner a generalized understanding of the workflow researchers
follow. We analyzed the 286 papers and identified three steps that are prevalent to data retrieval: (1) data sampling, which is the selection criteria for the data being analyzed, (2) data filtering, which is further filtering of the sampled data to remove irrelevant data, and (3) data retrieval, which is the collection of the data from its source.

    
    


\begin{figure}[htbp]
\centerline{\includegraphics[scale=.3]{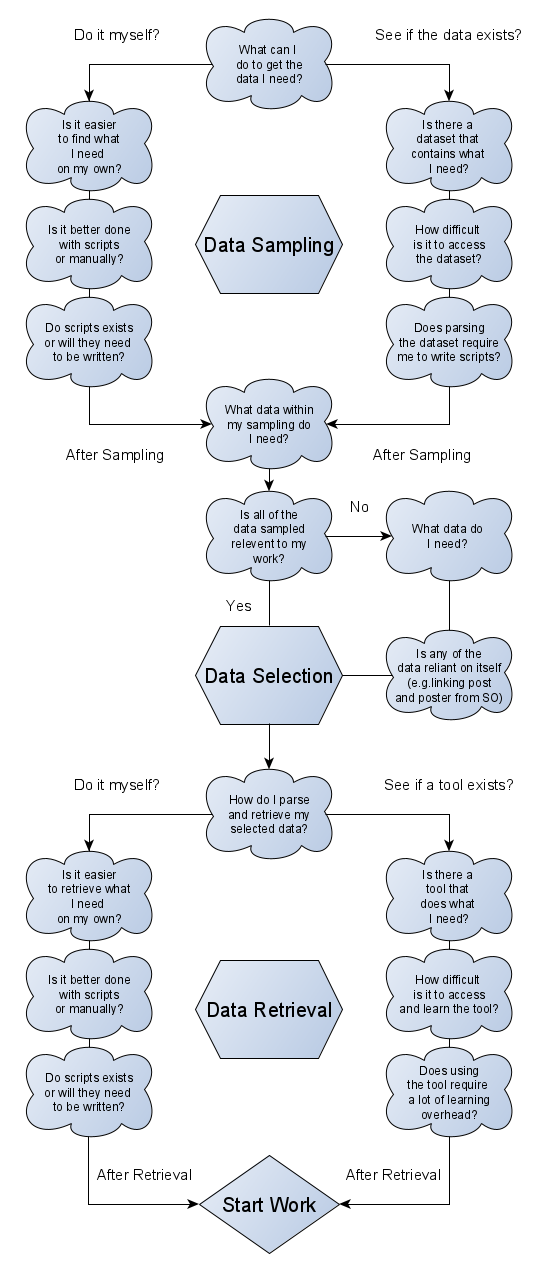}}
\vspace{-0.15in}
\caption{The generalized workflow that we inductively developed through our literature review.}
\label{fig-workflow}
\end{figure}

Figure~\ref{fig-workflow} illustrates this pattern and the steps possible for each section within, which we analyzed the possible paths for each step. It is important to note these steps were not always followed in the exact order listed and sometimes both paths for a step were taken. For example, if a researcher chose to use a dataset that had been previously compiled for mining software repositories, they would have been categorized as not doing the work themselves when sampling the data. However, sometimes researchers supplemented this dataset with more data they determined to be relevant. Thus, this would indicate they took both paths.

To help with understanding the process, we list three representative examples of workflows found in our literature review that exemplify the differences in conducting such research. The first workflow we present was performed by Alqaimi et al. in \cite{alqaimi2019automatically}. This team of researchers was interested in generating documentation for Lambda expressions in Java and chose to mine this data from GitHub. However, as they note in the paper, many of the repositories on GitHub are not related to software development. Thus, they chose to use the RepoReaper framework to sample projects that were non-trivial in nature. RepoReaper, developed by Munaiah et al. \cite{munaiah2017curating}, was engineered to help identify software projects. After sampling 51,392 Java projects via RepoReaper, they then cloned projects randomly in sets of 1000. They did this as a way to further filter the data until they retrieved 435 projects with lambda expressions. This required a total of 4000 projects cloned or 4 iterations of the random sampling. In the context of our workflow classification, this process would have been classified as starting with a pre-compiled dataset (the use of RepoReaper), performing further filtering, and doing the work themselves for the retrieval. Notably, in this process, the filtering process was combined with the retrieval process.

The second notable workflow was performed by Laaber et al. in \cite{laaber2018evaluation}. This team of researchers aimed to analyze the quality of microbenchmark suites by creating a Java and Go suite. They did this by first sampling all Java and Go projects on Google Big Query then mapping them back to their individual GitHub projects. Afterwards, they ordered them by stars and watchers and manually selected 5 of each language. Then, they manually checked the data for issues and re-used Google Big Query to cross reference how relevant these projects were. Thus, this paper's workflow would have been classified in our system as creating their own sampling criteria, performing a further filtering of the data, and retrieving the data manually/personally.

The final workflow we present was published by Menzies et al. in \cite{menzies2018500+}. This team of researchers set out to verify the work of previous deep learning in mining software repositories research. They did this by taking a pair of pre-compiled datasets and applying the data to a set of pre-made models. The results these models produced were very comparable to what was previously presented in another paper. Furthermore, the results were at least 500 times faster to produce. In our workflow analysis, this research would be classified as using someone elses data to perform the sampling, not performing any further filtering, and not having to do any form of retrieval.

As we kept track of the workflows, a typical process used by researchers became apparent. The overall data for this workflow is listed in Table~\ref{tab-workflows}. This data is based on what is/is not stated in the papers about how the researchers performed the work. We did not perform an in-depth analysis of external resources provided, such as replication packages. This means that the data may be unintentionally skewed based on a lack of reported details.

In the context of sampling, about 93.7\% or 268 of all of the papers collected stated how they performed their sampling. Out of the 268 papers that stated how they selected their initial set of data, 86.9\% performed a personal sampling of data for at least some of the work. Similarly, 23.1\% of the papers performed at least some of the work using a pre-compiled dataset. Note, these percentages do not add up to 100\% because as can be noted in Table~\ref{tab-workflows}, 25 papers performed a sampling using both methods. This data indicates that the vast majority of researchers choose to sample their data personally rather than rely on others work.

We also kept track of the number of papers that performed a further sampling/pruning of the data to retrieve only the data that was relevant. It is worth noting that papers that did not state one way or another were categorized as not performing this step. Thus, for the purpose of our review, all 268 of the papers that performed a sampling performed this step or did not. This step, termed filtering in our workflow diagram, was performed on 55.2\% of the papers reviewed. Thus, it can be inferred that researchers are not likely to collect the exact data they need on the first try. This could be caused by tools being too generalized across data domains.

The final step in the workflow is related to the actions performed to retrieve the data analyzed. Specifically, how the data collected was of interest. However, only about 63.6\% of the papers detailed how they performed the retrieval of data. This means that over a third of the papers analyzed in our literature review lacked sufficient information to reproduce the research. Note this is only in the context of what is stated within the paper. Many papers also provide supplemental material that may mitigate these issues. However, as papers are supposed to stand alone, this information is relevant and important to describe. The lack of such information affects the overall reproducibility of the work. 

There are many ways to perform the retrieval in mining software repositories research. As can be noted in Section~\ref{sec-results-tools}, there are many tools that researchers rely on to perform this work. However, many also create personal scripts/scrapers to perform their retrieval. Despite this fact, the information on the exact steps taken in these papers is lacking. For the papers that did provide this information, it can be determined that researchers were much less likely to perform the work themselves. Rather, they tended to rely on pre-built tools meant to help with the retrieval process. Of the 182 papers that detailed how they performed their retrieval, only about 33.5\% of them decided to do the work personally. Perhaps more interesting, 79\% of papers relied on others work, at least partially, to perform the scraping/mining process of the work. This once again does not add up to 100\% because 26 of the 186 papers used a tool for retrieval and retrieved the data personally.



\begin{table}[t]
\centering
\caption{This table describes the choices described in each paper for each step of their workflow. The total number of papers analyzed is 286. The filtering information only applies to papers that performed a sampling. (Percentages based on the total number of papers.)}
\vspace{-0.1in}
\begin{tabular}{|l|c|c|}
\hline
Workflow Choice & Sampling & Retrieval\\
\hline
Personal Work & 206 (72.0\%) & 35 (12.2\%) \\
Rely On Others Work & 35 (12.2\%) & 121 (42.3\%) \\
Both & 27 (9.4\%) & 26 (9.1\%) \\
Not Stated & 17 (5.9\%) & 94 (32.9\%) \\
Not Applicable & 1 (0.3\%) & 10 (3.5\%)  \\
\hline
\hline
Performed Filtering & Yes & No \\
\hline
 & 148 (51.7\%) & 120 (42.0\%) \\
\hline
\end{tabular}
\vspace{-0.15in}
\label{tab-workflows}
\end{table}


\subsection{RQ2 Results: Methodologies} \label{sec-results-methodologies}

\begin{table}[t]
\centering
\caption{Frequency of datasets from the 286 papers used in addition to GitHub or StackOverflow. Only datasets used more than once are included. (Percentages based on the total number of papers that used datasets)}
\vspace{-0.1in}
\begin{tabular}{|l|c|}
\hline
Dataset & Number of Papers Using \\
\hline
Presented in previous work & 34 (44.2\%) \\
RepoReaper & 8 (10.4\%) \\
F-Droid & 7  (9.1\%) \\
Jira & 7 (9.1\%) \\
Bugzilla & 5 (6.5\%) \\
CoinMarketCap.com & 3 (3.9\%) \\
CVD & 2 (2.6\%)\\
NVE & 2 (2.6\%) \\
\hline
\end{tabular}
\label{tab-datasets}
\vspace{-0.15in}
\end{table}

As mentioned in Section~\ref{sec-method}, how frequently datasets in mining software repositories research were used was of interest to our research. In pursuit of determining this, we had to define what we would categorize as a dataset. We categorized tools that allowed for an easy sampling of data or prebuilt datasets meant for direct re-use as datasets. Table~\ref{tab-datasets} shows the frequency of these datasets used in addition to either GitHub or StackOverflow (per our initial selection criteria for the literature review). All datasets that were used are listed on our spreadsheet located in our online appendix\footnote{https://zenodo.org/record/5274208\#.YSlMs7RKjs0}. Some papers use multiple of these datasets, though we excluded any dataset only used once. Interestingly, of 77 papers that used datasets, many of them were distributed in 2 main fields of interest. 

Specifically, papers tended to be interested in bug related datasets for software research. Of the 77 papers that used datasets 20.7\% of them were in this context. However, this is only when counted in combination. No single bug dataset was used in more than 7 papers. The datasets most frequently used in this context were Jira and BugZilla.
The other category of interest for researchers was related to characterizing non-trivial projects. The tool RepoReaper allows researchers to calculate the score of repositories based on best engineering practices. This practice has been used to identify engineered projects which many count as non-trivial. Similarly, F-Droid is a software repository for Android apps that only hosts free open-source applications. The site has links to where the code is hosted and researchers used this to identify non-toy applications in Android. Of the 77 papers that used these datasets, 19.5\% of the papers were focused on one of these two datasets. 

The remaining datasets fell into a range of categories and uses. Interestingly, while none of these datasets were used more than once, many of the datasets that were used were presented in previous papers. Categorizing their main use would be difficult without a deep dive into previous literature. However, 44.2\% of the datasets used were presented in previous works. It is possible a good portion of these datasets were reused in attempt to reproduce/replicate the previous works. Unfortunately, this relative lack of dataset reuse indicates that most often researchers would rather perform the data selection work personally. Thus, researchers that publish datasets should temper their expectations on reuse.


\begin{table}[t]
\centering
\caption{Forms of sampling performed in the 286 papers. (Percentages based on the entirety of papers)}
\vspace{-0.1in}
\begin{tabular}{|l|c|}
\hline
Criteria for Sampling & Times Used\\
\hline
Programming language & 172 (60.1\%) \\
Popularity & 125 (43.7\%)\\
Manual & 76 (26.6\%)\\
Sampling by dataset & 76 (26.6\%) \\
Library/package & 39 (13.6\%) \\
Randomly & 38 (13.3\%) \\
File inclusion/exclusion & 34 (11.9\%) \\
Forks & 27 (9.4\%)\\
Commits & 12 (4.2\%) \\
Lines of Code & 9 (3.1\%)\\
\hline
\end{tabular}
\label{tab-sampling}
\vspace{-0.15in}
\end{table}

Data sampling is a normal part of big data research. In traditional forms of research, researchers tend to perform the sampling in 1 of 5 ways. These sampling practices are categorized as random sampling, systematic sampling, sampling by convenience, clustered sampling, or stratified sampling. In mining software repositories research, sampling is also frequently performed due to the vastness of the data. However, performing the traditional forms of sampling can be difficult. Thus, we attempted to stratify what the most common forms of sampling were in this field of research. To keep track of this in the literature review, common forms of sampling were classified but not every form was included. Frequently, researchers needed to perform a very specific classification for their sampling that was unlikely to be used again. Thus, only the commonalities between papers were counted. 

According to our literature review, the forms of sampling tended to be performed based on the availability of data. For example, a researcher may classify data as relevant based on if the project/code is written in a certain coding language. Another common form is sampling by popularity. This is most frequently done in an attempt to weed out trivial projects in Git. This is frequently necessary because many projects hosted on GitHub are created as a toy project or in the context of academics. Sampling by popularity can be done in multiple ways, but is most often done by selecting projects or code by its rating. If a project is rated highly by others, it has less of a chance of being trivial in nature. The forms of sampling identified, listed in Table~\ref{tab-sampling}, were included due to their frequency of usage in the papers reviewed.

As can be noted, sampling based on language is the most popular form of sampling in mining software repositories research. Out of the 265 papers that stated how they performed their sampling, 64.2\% of them included language(s) in the criterion. Similarly, sampling based on popularity was a metric frequently used with 45.3\% of the papers using this as a selection factor. Interestingly, 29.7\% of papers performed a sampling manually. The manual sampling was often performed after sampling based on some of these other criterion. As can be noted in Table~\ref{tab-sampling}, the forms of sampling are very heuristic based. Interestingly, only one of the traditional forms of sampling, random sampling, was performed. Alongside this, the random sampling practices did not follow a consistent format. Indeed, no two papers performed their random sampling in the same way. Alongside this, many of the papers performed another sampling first based on these other heuristic based restrictions then randomly sampled that result. Furthermore, many of the papers gave no insight into how exactly they randomly sampled their data and rather opted to solely state it was done.

Of the papers that did perform a random sampling, a few stick out as interesting and excellent examples. For example, Treude et al. \cite{treude2019predicting} performed a random sampling of GitHub projects. To do this, they wrote a script to generate a number between 0 and the total number of projects hosted on GitHub. After finding the project associated with that number, they checked if it contained code from a set of 8 coding languages. If it did, they also checked if the README contained at least 100 non-ASCII characters. Satisfying these conditions, the project was included in their sample. They did this until they had a total of 5000 projects. This required them to write personal scripts to perform the work, because there is no easy way currently to such work on GitHub. 

Another example of a random sampling comes from Barik et al. \cite{barik2018should}. They performed a sampling upon Stack Overflow questions specifically focused on questions about compiler errors in 25 languages. They first stratified the questions according to language then performed a simple-random sampling on each stratum in which question-answer pair was equally possible to be selected. Then they checked if these pairs fit their criterion (i.e. not mis-labelled, not trolling, etc). If the pair fit, they included it until they had 30 pairs from each of the top 7 languages, for a total of 210 pairs to analyze.

Unfortunately, good examples of this form of sampling are fairly rare. Many of the papers analyzed that performed a random sampling did so on data they had already performed another sampling on. Alongside this, they frequently gave no insight into how the random sampling was performed. Such non-standard forms of random sampling could obscure the steps to reproducing works of research. Thus, analyzing and bettering our understanding on this practice is integral to continuing improvement of our research field.



\subsection{RQ3 Results: Tools} \label{sec-results-tools}

In mining software repositories research, mining of the data is an essential step in most researchers workflows. Thus, tool creation to help ease this process is important. Stratifying what tools are used can give insight into the mining software repositories process. To gather this information, we kept track of the mention of any tools used for general retrieval of data. An example of such a tool is the GitHub REST API. Overall, these tools play a major role in how researchers perform their work. 

\begin{table}[t]
\centering
\caption{Frequency of tools used as part of the mining software repositories workflow in the 268 papers reviewed. We excluded any tool used only once. (Percentages based on the total number of papers that used tools)}
\vspace{-0.1in}
\begin{tabular}{|l|c|}
\hline
Form of Sampling & Times Used\\
\hline
GitHub REST APIs & 68 (44.7\%) \\
GHTorrent & 33 (21.7\%)\\
Stack Overflow Dump & 22 (14.5\%) \\
SOTorrent & 17 (11.2\%) \\
Google Big Query & 10 (6.6\%)\\
Software Heritage & 5 (3.3\%)\\
Boa & 5 (3.3\%)\\
Travis CI API & 5 (3.3\%)\\
PyDriller & 4 (2.6\%)\\
World of Code & 3 (2.0\%)\\
Perceval & 3 (2.0\%)\\
ScraPy & 2 (1.3\%)\\
\hline
\end{tabular}
\label{tab-tools}
\vspace{-0.15in}
\end{table}

Despite how numerous and available tools seem to be, the tool usage is very skewed towards just a few of the tools in the field.
It is worth noting, there were a total of 152 papers that stated they used tools. However, the total number of tools used in Table~\ref{tab-tools} equates to more than 152. Alongside this, the table only lists tools that were used more than once in the research. This data disparity was caused by papers that used multiple tools. Thus, tool statistics are only relevant to the tool it describes. 

In our literature review, the tool used most frequently is the GitHub REST API. Of the 152 papers that stated usage of tooling, 44.7\% of the papers used the GitHub REST APIs for at least a portion of the work. GitHub has a built in REST API that allows for data retrieval and easier sampling of projects. As can be noted in Table~\ref{tab-tools}, this tool is used more than twice as frequently as the next tool. It's frequent usage indicates how convenient researchers find the tool and also indicates the reliance on GitHub specific data.

The next most used tool, GHTorrent, is a tool built to mirror the GitHub timeline and provide a database to easily access the data. The tool, designed by Gousi et al. \cite{GHtorrentGousi13}, was created for research in software analytics. Since it was published, the tool has quickly grown in popularity and since its data source is GitHub, it provides a lot of data to analyze. Of the 152 papers that provided information into their retrieval process, 21.7\% of them used GHTorrent for at least a portion of their work. Similar to the GitHub REST APIs, this tool is heavily relied on since it provides a platform that allows researchers to perform most of their work. Performing the sampling and retrieval steps is easily done through this tool.

The next two most frequently tools are specific to Stack Overflow data. The Stack Overflow Data Dump and the SOTorrent dataset are frequently used when performing code snippet research. The Stack Overflow Dump is a compilation of anonymous data published quarterly from Stack Exchange. The SOTorrent \cite{baltes2018sotorrent} dataset mimics the purpose of GHTorrent but draws its data from the official Stack Overflow data dump. Of the 152 papers that utilized tools for some of the work, 14.5\% of them used the official data dump and 11.2\% used the SOTorrent dataset.

None of the rest of the tools were used by more than 10\% of the papers. However, some of them standout as excellent tools for the work despite their relative lack of prevalence. Tools like Software Heritage, Boa, PyDriller, and the World of Code address many needs for performing mining software repositories research, however, they seem to lack usage in our literature review. It can be gathered that mining software repositories research is primarily reliant on the tools previously listed. 

Indeed, considering the primary reliance on these tools, we also wanted to determine if there is a correlation between the usage of these tools. Interestingly, there was little correlation between the tools being used. The tools used most often together were a combination of GHTorrent and the GitHub REST APIs. As can be noted in table \ref{tab-tools}, a total of 33 papers used GHTorrent. Of these papers, a total of 13 of those used the GitHub REST APIs as well. GHTorrent provides a link to the actual GitHub project being analyzed and thus, if researchers needed additional information, it would be easily accessed through the REST APIs. 

In a similar vein, tools like the Travis CI API had a correlation to the GitHub REST APIs as well. Of the 5 papers that used Travis, 3 of them also used the GitHub APIs. PyDriller is similarly correlated with 4 papers using the tool and 2 of those also using the GitHub APIs. Unfortunately, such a small sample size makes it hard to garner a clear and certain correlation between these tools. The GHTorrent dataset is built from the GitHub ecosystem and thus inherently has some correlation. The other tools do not necessarily correlate at such a meta level.


As we were interested in the tooling being used, we were also interested in determining how frequently researchers wrote scripts to perform their work. To determine this, we attempted to stratify the script writing information into the three steps described in the generalized workflow of researchers. Specifically, we were interested in determining which step in the research workflow was most likely to motivate researchers to perform the work personally. 

\begin{table}[t]
\centering
\caption{The frequency of code scripts written, not written, or not stated during the three major steps of mining software repositories in the 286 papers reviewed. (Percentages based on the total number of papers.)}
\vspace{-0.1in}
\begin{tabular}{|l|c|c|}
\hline
Step & Scripts & Number of Times \\
\hline
 & Written & 69 (24.1\%) \\
Data Sampling & Not Written & 100 (35.0\%)\\
 & Not Stated & 117 (41.0\%)\\
\hline
 & Written & 78 (27.3\%)\\
Data Filtering & Not Written & 87 (30.4\%)\\
 & Not Stated & 121 (42.3\%)\\
\hline
 & Written & 76 (26.6\%) \\
Data Retrieval & Not Written & 67 (23.4\%)\\
 & Not Stated & 143 (50.0\%) \\
\hline
\end{tabular}
\label{tab-scripts}
\vspace{-0.15in}
\end{table}

Our analysis on this topic sometimes required answers to be inferred from the vastness/or lack thereof of data. For example, many papers use tools such as the Stack Overflow Data Dump to get the data they plan to work with. However, they frequently mention the need to further classify or restrict that data in some way. Given the size of the data available in the Stack Overflow Data Dump, it can be reasonably inferred that classification scripts would be necessary for this work. 
Making this connection is necessary to garner a more complete look at the script making process in repository mining. This is because many papers do not directly state they wrote scripts. However, even with this connection, the content in the papers could not always be classified in yes or no fashion. If the answer for script making could not be reasonably inferred in either direction, the paper was categorized as unclear. 

As can be noted in Table \ref{tab-scripts}, the information on script writing was not exceptionally well described. As can be noted in Table \ref{tab-replication}, 145 of the papers that published replication kits included personal scripts of some kind. However, no more than 78 of the papers reviewed noted the specific use of these scripts. Thus, it's clear that such potentially relevant information is being excluded in descriptions of researchers work. Since our review into the replication kits does not attempt to determine the use of scripts, it is possible they were not used to sample, filter, or retrieve data for the researchers. However, the lack of descriptions may be problematic for reproducibility. 

It is clear from the information provided that researchers most frequently wrote scripts when filtering/pruning the data they originally sampled. However, a close second comes from the researchers that chose to implement personal scripts of some sort to collect/retrieve their desired data. Interestingly, researchers were least likely to write scripts to sample their data. This could indicate the available tooling in this field is generally sufficient to fit researchers needs. Or perhaps, that writing scripts on this specific step is more difficult than for the others. However, without more descriptive information on the topic, garnering a good understanding of this will be difficult.



\subsection{RQ4 Results: Reproducibility} \label{sec-results-replication}



\begin{table}[t]
\centering
\caption{This table shows the general replication package information from the 286 papers reviewed. N/A means that the paper did not publish a dataset or replication kit.}
\vspace{-0.1in}
\begin{tabular}{|l|c|c|c|}
\hline
 & True & False & N/A\\
\hline
Paper states data is available & 188 (65.7\%) & 29 & 69 \\
Paper states scripts are available & 122 (42.7\%) & 95 & 69 \\
Paper states other is available & 86 (30.0\%) & 131  & 69 \\
Replication package link works & 205 (71.7\%) & 12 & 69 \\
Data is available & 183 (64.0\%) & 23 & 80  \\
Scripts are available & 145 (50.7\%) & 60 & 80 \\
Other resources are available & 115 (40.2\%) & 90 & 80\\
\hline
\end{tabular}
\label{tab-replication}
\vspace{-0.15in}
\end{table}

In pursuit of answering RQ4, information related to researchers replication packages was tracked to garner insight into the overall reproducibility of the work. As there has been some confusion on the definition of replicability and reproducibility in recent years, we chose to base our work off the ACM definition listed on their official website ~\cite{ACM2020Reproducible}. Specifically, we define reproducibility as the ability for a new team of researchers to perform the same experiment with the same experimental setup and garner the same results. Replicability is defined as a new team being able to garner the same results with a differing experimental setup. This is how we will define our research into this topic going forward.
%

To do this, we analyzed first the direct statements made on replication packages for each paper then the packages themselves. This was done in yes/no fashion to help avoid bias. In order to perform this research, we identified 3 main factors of replication packages that we believe might be useful for reproducibility. Specifically, we tracked if the papers stated the final dataset they used was available, if scripts they used were available, and if other resources outside this scope (i.e. machine learning models) were available. Then, we navigated to the outside resources (when possible), and verified if at a glance this information seemed to be available. It is necessary to note, we did not do an in-depth analysis of these packages. Thus, if researchers provided scripts in these packages, we did not analyze what the expected use was. We simply aimed to verify the veracity of the claims made in a quick and simple way.

We provide the statistics for this section in two steps. The first step is solely related to what the researchers stated about their replication package. This provides insight into how researchers communicate their resources to the general readership. In the second step, we provide corollary data between the two. It is worth noting that not every paper analyzed provided a replication package or labelled it as such. For our purposes, we decided the definition of a replication package solely relates to providing resources that may help with replicating the work. Thus, if researchers state they make their dataset public for future analysis, but lack any other information on the process, we still count that as publishing a replication package. This caveat applies to each of the possible steps in the replication package. For our analysis, 217 of the 286 papers analyzed, or 75.9\%, fit this criteria and were counted as publishing a replication package. The general statistics on these packages can be seen in Table~\ref{tab-replication}. 

As can be noted in our table, of the 217 papers that published a replication package 86.6\% of them claimed to have published a dataset for analysis. Interestingly, this was the only form of information that was shared consistently. Only 56.2\% of the papers shared if there were scripts contained in the package. Alongside this, only 39.6\% of the papers shared if their results/other resources of their analysis were available in the replication package. This information can be related to the actual product it provided in interesting ways. First it is worth noting that 12 papers had non-working links to their packages. Thus we ignore those papers for the statistics on what was contained within the packages. Unsurprisingly, of the 205 papers that did have working packages, 89.3\% of them provided datasets. This is a comparable statistic to what the papers stated. However, interestingly, the scripts and other resources are much more likely to exist as opposed to what was stated in the papers. Scripts existed in 70.7\% of the packages and 56.1\% of the packages included additional resources (e.g., Docker containers).

Thus, it can be seen that the general information on these replication packages is lacking. The information/data contained within them is not well defined and this could make use of such replication packages difficult/impossible.


\section{Discussion} \label{sec-discussion}

In summary our findings provide some interesting insight into the field of mining software repositories research. From our review, we found that the way researchers report their workflow lacks a standard format. While most of the papers analyzed included a section on their methodology, the exact format of these sections varied greatly. The sections frequently lack relevant information or glosses over the exact steps taken to perform their work. In section \ref{sec-discussion-meth}, we discuss the implications of this and provide a suggested format for improving the reporting of these methodologies. Alongside this, we discovered that researchers in the field frequently rely on a list of tools and datasets to perform their work. They also frequently create and publish new datasets or tools in the process of performing their research. However, the vast majority of tool and dataset usage seems to be restricted to a small subset of the total tools available. Frequently, researchers rely on the very short tail of the available tool distribution to perform their work. We analyze and discuss these findings in sections \ref{sec-discussion-reuse} and \ref{sec-discussion-tools}. 

\subsection{Standard for Reporting Methodologies} \label{sec-discussion-meth}
In the process of our literature review we analyzed the methodological workflow researchers followed when performing their work. Specifically we were interested in the reproducibility of the work they performed. In the process of this review, we realized there is a disparity in reporting methodologies between papers. While almost every paper had a section reporting their methodology, many of these lacked critical information on steps performed. As noted in Section~\ref{sec-results-methodologies}, many papers lacked information on how researchers performed their data retrieval. Frequently, the steps were obscure and what tooling they relied on to perform the work was never directly stated. While it is possible to perform mining software repositories research without the reliance on tools, such a process should be indicated by the team. However, a total of 94 of the 286 papers analyzed gave no insight at all into how the data was retrieved (this data can be referenced in table \ref{tab-workflows} in Section \ref{sec-results-workflows}). Such a lack of information, based on the ACM definition of reproducible, would indicate that the work is at least partially not reproducible because it lacks information on the experimental setup. 

In a similar vein of focus, many research papers provided some form of sampling criterion. However, such criterion often lacked critical information such as the timestamp at which the data was selected. Papers frequently choose to sample data based on popularity or other ever changing facets of open source software. Thus, a project that may rank as one of the most popular projects in the field in one year could lack similar levels of use the following year. Thus, if researchers were interested in replicating the work, performing the same steps to sample the data might lead to different projects/data sampled. Unfortunately, this timestamp information is completely missing from 175 of the 286 papers analyzed. This issue would indicate the works are not reproducible when based solely upon what was stated in the papers. 

These points on reproducibility might be argued because, as we note in our review, many of these papers published a replication package to assist with reproduction of their work. Indeed 217 of the 286 papers did provide some form of replication package. It is worth noting that of the 94 papers that gave no information on retrieval, 65 of them provide a replication package of some form. Similarly, of the 175 papers that gave no timestamp information on when they collected the data, 122 of them contained some form of replication package. Thus, this judgement of non-reproducibility can be argued.

However, academic research papers are often expected to make the work stand alone without the need of outside artifacts. Such a demand indicates that, regardless of the supplementary material provided, future works should strive to fulfill these conditions to the best of their abilities. In hopes of assisting future research in reproducibility, we set out to categorize the information most relevant to reproducibility of a work. To do this, we analyzed which papers provided less than adequate information to allow for reproduction. 

To assist with this endeavor, we provide a suggested standardized format to help future researchers with making sure their work remains reproducible. As research in this field is constantly changing, we do not claim this to be an comprehensive list. However, as indicated by our literature review, this information is relevant to a vast majority of mining software repositories research. It is up to the individual researcher teams to note missing information applicable to themselves. We break this information into the three steps considered in the workflow diagrams.

For sampling data these metrics are as follows:

\begin{itemize}
    \item From what source was the data sampled? (i.e. the website/code host the data was sampled)
    \item Were any personal scripts written or tools used for this step and what do they do?
    \item What was the total list of sampling criterion and why?
    \item When was the data sampled?
\end{itemize}

For filtering data these metrics are as follows:

\begin{itemize}
    \item Was a filtering of the original sample performed?
    \item What was the filtering criterion?
    \item How was it done (i.e. manually, through scripts, etc)?
\end{itemize}

For retrieval of data these metrics are as follows:

\begin{itemize}
    \item How was the data retrieved (i.e. through personal effort, reliance on tools, etc)?
    \item When was the data retrieved (Not always done at the same time as the sample)?
    \item If tools were relied on, did it provide all of the information needed?
    \item Were any scripts written for this step?
\end{itemize}

It is worth noting that not all of this information will always be pertinent to provide. Some researchers may choose not to perform any filtering on their first sample of data. Others may not perform any retrieval at all and rely instead on a dataset previously created. Thus, it is up to the specific research team to determine how much of this format is relevant to their work.

\subsection{Reuse of Data and Tools} \label{sec-discussion-reuse}
One notable goal of our literature review was to determine the tools used in the research. From this research it became clear that not many tools tended to be used. Indeed, researchers tended to rely on a small sample of tools for almost all of the work performed. Indeed, as can be noted in section \ref{sec-results-tools}, the use of tools can be attributed to a scant few of the possible options. Within these popular tools, not all of them necessarily were created with research in mind. Indeed, tools like GHTorrent and SOTorrent were specifically created for this purpose. However, tools like the GitHub REST APIs are less directly focused on such goals. This is notable because other tools specifically designed for research purposes seem to suffer in terms of usage. Most likely this is due to the difficulty of creating a tool with access to as much relevant and interesting data as the GitHub REST APIs. The data storage of such information would certainly be enormous and accessing such information needs to be easily done. Some tools have attempted such things and they tend to be the tools relied on most in the research. Tools like GHTorrent and SOTorrent contain vast amounts of data in an easily accessed format and thus get used frequently. This could indicate that future tools in the field may be better served by copying the model of these tools. Tools like Boa and the World of Code provide similar resources and as they grow their frequency of use does too. 

Datasets on the other hand are not so easily categorized for the research. Many of datasets were noted in our literature review, but many of them were only used once. These datasets were very frequently presented in past research and often the use was in attempt to recreate or replicate that work. Thus, it can be seen that datasets do get used when they are published, but it is often not for any other reason than to verify previous work. Thus, researchers publishing data they desire for others to build upon may be disappointed. However, there are some notable exceptions to this. As noted in section \ref{sec-results-methodologies}, the datasets most frequently reused were ones that provided a large sample of data that could be categorized as non-trivial in nature or data that required much manual effort to curate (such as bug datasets). These datasets are unique in that they provide data that is otherwise not easily accessible. 


\subsection{Implications for Tools} \label{sec-discussion-tools}
Sampling is a common action in most legitimate sciences as scientists frequently are faced with enormous amounts of data. Sampling is usually grouped into two categories based on if the sampling is based on a probability or not. Non-probability sampling tends to be done for convenience to help with creating hypotheses. However, for a more systematic analysis it is often better to perform a sampling based on probability. There are a number of ways to perform this form of sampling including simple random sampling, stratified, systematic, and cluster sampling. In other fields, random sampling is frequently chosen to reduce bias. However, in the context of mining software repositories research, this form of sampling is rarely ever utilized. In fact, of the 287 papers analyzed only 38 papers performed a random sampling of any sort. Alongside this, frequently this random sampling was done after an initial sampling of some sort. The reason for this seems to be attributed to the difficultly of performing such an action. The tools readily available to the public have no built in tooling to assist with this process. 

Indeed, performing a true random sampling of data is very difficult to do. Of the papers performing a random sampling, very few seemed to do so in truly random fashion. Furthermore, it required much manual effort on their part to do. Researchers in \cite{treude2019predicting} performed a random sampling of projects on GitHub by randomly generating a number between one and the total number of projects hosted on GitHub then checked if they fit the criteria they desired. This involved verifying the project randomly selected contained code from at least one of eight programming languages and if the README contained more than 100 non-ASCII characters. Such intensive work required the researchers to write scripts just to perform their sample and indicates the general difficulty researchers face when attempting to sample randomly. 

Considering the difficulty of performing this work, it is apparent why researchers tend to avoid this method despite its prevalence in other fields. Indeed, many researchers rely on a heuristic based sampling based on popularity or specific characteristics of a project. However, there are potential solutions to this challenge. While most of the mining software repositories tools provide no assistance with this form of sampling, one tool does allow for random sampling. The World of Code is a tool that attempts to document the entire FLOSS ecosystem by pulling its data from the large software repository hosting sites. This data is then stored in a set of databases that researchers can access. 
%
It provides a mechanism for researchers to perform a random sampling easily. 

\subsection{Threats to Validity} \label{sec-threats}

Our literature review, much like others, suffers from the standard threats to validity in the field. There is some risk of selection bias being introduced due to the process by which we chose relevant papers. However, we attempted to mitigate these issues by analyzing three different conferences across the software engineering domain. There is also some risk of historical bias because the data was selected over a set of years. However, we believe this bias is appropriately addressed by the relatively short range of years analyzed. Finally, there is some risk of human error as portions of our data were categorized manually. 

\section{Conclusions} \label{sec-conclusion}

In this paper, we have contributed the findings from a systematic literature review of mining software repositories.
In particular, we analyzed 286 papers from recent conference proceedings in regards to the workflows, methodologies, reproducibility, and tools used by researchers to understand the state-of-the-art and to highlight potential challenges.
A few key findings of our work include:

\begin{itemize}
    \item Workflows are disjoint, and often require using multiple tools and custom scripts.
    \item Papers often do not provide enough methodological information to be reproducibile. For example, 33\% of papers provide no information on how the data was retrieved.
    \item Datasets published by researchers are rarely used by others.
\end{itemize}

Furthermore, we contributed ways to improve existing tools and also provide recommendations to improve research workflows and the reproducibility of research through a standard for reporting methodologies.  
Understanding the state-of-the-art and working towards a standard for reporting research is of paramount concern to mitigate a "replication crisis" in software engineering research.

\balance

%

\bibliographystyle{plain}
\bibliography{main.bib}

\begin{thebibliography}{10}

\bibitem{alqaimi2019automatically}
Anwar Alqaimi, Patanamon Thongtanunam, and Christoph Treude.
\newblock Automatically generating documentation for lambda expressions in
  java.
\newblock In {\em 2019 IEEE/ACM 16th International Conference on Mining
  Software Repositories (MSR)}, pages 310--320. IEEE, 2019.

\bibitem{DBLP:conf/msr/BaltesDT008}
Sebastian Baltes, Lorik Dumani, Christoph Treude, and Stephan Diehl.
\newblock Sotorrent: reconstructing and analyzing the evolution of stack
  overflow posts.
\newblock In Andy Zaidman, Yasutaka Kamei, and Emily Hill, editors, {\em
  Proceedings of the 15th International Conference on Mining Software
  Repositories, {MSR} 2018, Gothenburg, Sweden, May 28-29, 2018}, pages
  319--330. {ACM}, 2018.

\bibitem{baltes2018sotorrent}
Sebastian Baltes, Lorik Dumani, Christoph Treude, and Stephan Diehl.
\newblock Sotorrent: Reconstructing and analyzing the evolution of stack
  overflow posts.
\newblock In {\em Proceedings of the 15th international conference on mining
  software repositories}, pages 319--330, 2018.

\bibitem{barik2018should}
Titus Barik, Denae Ford, Emerson Murphy-Hill, and Chris Parnin.
\newblock How should compilers explain problems to developers?
\newblock In {\em Proceedings of the 2018 26th ACM Joint Meeting on European
  Software Engineering Conference and Symposium on the Foundations of Software
  Engineering}, pages 633--643, 2018.

\bibitem{msrgit}
Christian Bird, Peter~C. Rigby, Earl~T. Barr, David~J. Hamilton, Daniel~M.
  German, and Prem Devanbu.
\newblock The promises and perils of mining git.
\newblock {\em 2013 10th Working Conference on Mining Software Repositories
  (MSR)}, 0:1--10, 2009.

\bibitem{dabic2021sampling}
Ozren Dabic, Emad Aghajani, and Gabriele Bavota.
\newblock Sampling projects in github for msr studies.
\newblock {\em arXiv preprint arXiv:2103.04682}, 2021.

\bibitem{BOAdyer2015}
Robert Dyer, Hoan~Anh Nguyen, Hridesh Rajan, and Tien~N Nguyen.
\newblock Boa: Ultra-large-scale software repository and source-code mining.
\newblock {\em ACM Transactions on Software Engineering and Methodology
  (TOSEM)}, 25(1):1--34, 2015.

\bibitem{flint2021escaping}
Samuel~W Flint, Jigyasa Chauhan, and Robert Dyer.
\newblock Escaping the time pit: Pitfalls and guidelines for using time-based
  git data.
\newblock {\em arXiv preprint arXiv:2103.11339}, 2021.

\bibitem{GM03}
Daniel German and Audris Mockus.
\newblock Automating the measurement of open source projects.
\newblock In {\em ICSE '03 Workshop on Open Source Software Engineering}, page
  Automating the Measurement of Open Source Projects, Portland, Oregon, May
  3-10 2003.

\bibitem{GHtorrentGousi13}
Georgios Gousios.
\newblock The ghtorrent dataset and tool suite.
\newblock In {\em Proceedings of the 10th Working Conference on Mining Software
  Repositories}, MSR '13, pages 233--236, Piscataway, NJ, USA, 2013. IEEE
  Press.

\bibitem{Kagdi2007JSME}
Huzefa Kagdi, Michael~L. Collard, and Jonathan~I. Maletic.
\newblock A survey and taxonomy of approaches for mining software repositories
  in the context of software evolution.
\newblock {\em Journal of Software Maintenance and Evolution: Research and
  Practice}, 19(2):77--131, 2007.

\bibitem{kitchenham2009systematic}
Barbara Kitchenham, O~Pearl Brereton, David Budgen, Mark Turner, John Bailey,
  and Stephen Linkman.
\newblock Systematic literature reviews in software engineering--a systematic
  literature review.
\newblock {\em Information and software technology}, 51(1):7--15, 2009.

\bibitem{laaber2018evaluation}
Christoph Laaber and Philipp Leitner.
\newblock An evaluation of open-source software microbenchmark suites for
  continuous performance assessment.
\newblock In {\em 2018 IEEE/ACM 15th International Conference on Mining
  Software Repositories (MSR)}, pages 119--130. IEEE, 2018.

\bibitem{WOCma2019world}
Yuxing Ma, Chris Bogart, Sadika Amreen, Russell Zaretzki, and Audris Mockus.
\newblock World of code: an infrastructure for mining the universe of open
  source vcs data.
\newblock In {\em 2019 IEEE/ACM 16th International Conference on Mining
  Software Repositories (MSR)}, pages 143--154. IEEE, 2019.

\bibitem{menzies2018500+}
Tim Menzies, Suvodeep Majumder, Nikhila Balaji, Katie Brey, and Wei Fu.
\newblock 500+ times faster than deep learning:(a case study exploring faster
  methods for text mining stackoverflow).
\newblock In {\em 2018 IEEE/ACM 15th International Conference on Mining
  Software Repositories (MSR)}, pages 554--563. IEEE, 2018.

\bibitem{MFH00}
A.~Mockus, R.~F. Fielding, and J.~Herbsleb.
\newblock A case study of open source development: The {A}pache server.
\newblock In {\em 22nd International Conference on Software Engineering}, pages
  263--272, Limerick, Ireland, June 4-11 2000.

\bibitem{M02}
Audris Mockus.
\newblock Measurement in software projects: taking advantage of version control
  repositories.
\newblock In {\em International Software Engineering Network, 2002}, Nara,
  Japan, October 2002.

\bibitem{MTutorial06}
Audris Mockus.
\newblock How to run empirical studies using project repositories.
\newblock 4th International Advanced School of Empirical Software Engineering,
  September 20, 2006, Rio de Janeiro, Brazil 2006.
\newblock Tutorial.

\bibitem{Changes07}
Audris Mockus.
\newblock Software support tools and experimental work.
\newblock In V~Basili and et~al, editors, {\em Empirical Software Engineering
  Issues: Critical Assessments and Future Directions}, volume LNCS 4336, pages
  91--99. Springer, 2007.

\bibitem{M14}
Audris Mockus.
\newblock Engineering big data solutions.
\newblock In {\em ICSE'14 FOSE}, 2014.

\bibitem{Mockus00}
Audris Mockus and Lawrence~G. Votta.
\newblock Identifying reasons for software change using historic databases.
\newblock In {\em International Conference on Software Maintenance}, pages
  120--130, San Jose, California, October 11-14 2000.

\bibitem{munaiah2017curating}
Nuthan Munaiah, Steven Kroh, Craig Cabrey, and Meiyappan Nagappan.
\newblock Curating github for engineered software projects.
\newblock {\em Empirical Software Engineering}, 22(6):3219--3253, 2017.

\bibitem{ACM2020Reproducible}
none.
\newblock Artifact review and badging - current, Aug 2020.

\bibitem{PyDrillerSpadini2018}
Davide Spadini, Maur\'{i}cio Aniche, and Alberto Bacchelli.
\newblock {PyDriller: Python framework for mining software repositories}.
\newblock In {\em Proceedings of the 2018 26th ACM Joint Meeting on European
  Software Engineering Conference and Symposium on the Foundations of Software
  Engineering - ESEC/FSE 2018}, pages 908--911, New York, New York, USA, 2018.
  ACM Press.

\bibitem{treude2019predicting}
Christoph Treude and Markus Wagner.
\newblock Predicting good configurations for github and stack overflow topic
  models.
\newblock In {\em 2019 IEEE/ACM 16th International Conference on Mining
  Software Repositories (MSR)}, pages 84--95. IEEE, 2019.

\bibitem{tu2018careful}
Feifei Tu, Jiaxin Zhu, Qimu Zheng, and Minghui Zhou.
\newblock Be careful of when: an empirical study on time-related misuse of
  issue tracking data.
\newblock In {\em Proceedings of the 2018 26th ACM Joint Meeting on European
  Software Engineering Conference and Symposium on the Foundations of Software
  Engineering}, pages 307--318, 2018.

\bibitem{Zimmermann}
Thomas Zimmermann, Peter Weissgerberv, Stephan Diehl, and Andreas Zeller.
\newblock Mining version histories to guide software changes.
\newblock {\em IEEE Transactions of Software Engineering}, 30(9), 2004.

\end{thebibliography}


\end{document}